\documentclass[unnumsec,webpdf,contemporary,large]{oup-authoring-template}%

\graphicspath{{Fig/}}

\theoremstyle{thmstyleone}%
\theoremstyle{thmstyletwo}%
\theoremstyle{thmstylethree}%

\begin{document}

\journaltitle{Bioinformatics}
\DOI{DOI added during production}
\copyrightyear{YEAR}
\pubyear{YEAR}
\vol{XX}
\issue{x}
\access{Published: Date added during production}
\appnotes{Original Paper}

\firstpage{1}


\title[]{Empowering Chemical Structures with Biological Insights for Scalable Phenotypic Virtual Screening}
\author[1]{Xiaoqing Lian}
\author[1]{Pengsen Ma}
\author[1]{Tengfeng Ma}
\author[1]{Zhonghao Ren}
\author[1]{Xibao Cai}
\author[1]{Zhixiang Cheng}
\author[1]{Bosheng Song}
\author[2]{He Wang}
\author[2]{Xiang Pan}
\author[3,$\ast$]{Yangyang Chen}
\author[4,$\ast$]{Sisi Yuan}
\author[5,$\ast$]{Chen Lin}

\address[1]{\orgdiv{State Key Laboratory of Chemo and Biosensing, College of Computer Science and Electronic Engineering; The Ministry of Education Key Laboratory of Fusion Computing of Supercomputing and Artificial Intelligence}, \orgname{Hunan University}, \orgaddress{410082, Changsha, China}}
\address[2]{\orgdiv{School of Artificial Intelligence and Computer Science}, \orgname{Jiangnan University}, \orgaddress{214122, Wuxi, China}}
\address[3]{\orgdiv{Department of Computer Science}, \orgname{University of Tsukuba}, \orgaddress{Tsukuba, Ibaraki, Japan}}
\address[4]{\orgdiv{School of Chinese Medicine}, \orgname{Hong Kong Baptist University}, \orgaddress{Kowloon Tong, Kowloon, 999077, Hong Kong SAR, China}}
\address[5]{\orgdiv{School of Informatics, National Institute for Data Science in Health and Medicine}, \orgname{Xiamen University}, \orgaddress{361000, Xiamen, China}}

\corresp[$\ast$]{Corresponding author. \href{mailto:chen.yangyang.xp@alumni.tsukuba.ac.jp}{chen.yangyang.xp@alumni.tsukuba.ac.jp} }
\corresp{ Sisi Yuan. \href{mailto:sisiyuan@hkbu.edu.hk}{sisiyuan@hkbu.edu.hk} }
\corresp{ Chen Lin. \href{mailto:cheyenne.lin@foxmail.com}{cheyenne.lin@foxmail.com}}



\abstract{
Motivation: The scalable identification of bioactive compounds is essential for contemporary drug discovery. This process faces a key trade-off: structural screening offers scalability but lacks biological context, whereas high-content phenotypic profiling provides deep biological insights but is resource-intensive. The primary challenge is to extract robust biological signals from noisy data and encode them into representations that do not require biological data at inference.\\
Results: This study presents DECODE (DEcomposing Cellular Observations of Drug Effects), a framework that bridges this gap by empowering chemical representations with intrinsic biological semantics to enable structure-based in silico biological profiling. DECODE leverages limited paired transcriptomic and morphological data as supervisory signals during training, enabling the extraction of a measurement-invariant biological fingerprint from chemical structures and explicit filtering of experimental noise. Our evaluations demonstrate that DECODE retrieves functionally similar drugs in zero-shot settings with over 20\% relative improvement over chemical baselines in mechanism-of-action (MOA) prediction. Furthermore, the framework achieves a 6-fold increase in hit rates for novel anti-cancer agents during external validation.\\
Availability and implementation: The codes and datasets of DECODE are available at \url{https://github.com/lian-xiao/DECODE}.
}

\maketitle

\section{Introduction}
Early-stage drug discovery aims to identify novel therapeutic candidates within a nearly infinite chemical space~\cite{goh2017deep,adil2021single,swinney2011were}. Success depends not only on molecular binding affinity but also on the ability of compounds to induce desired functional perturbations in complex biological systems~\cite{swinney2013contribution,lian2024advancing,lian2025inductive}. Consequently, phenotypic evaluation of large compound libraries is essential for prioritizing high-quality leads and reducing the high attrition rates typically seen in clinical development~\cite{hughes2021high,haasen2017phenotypic}.

Despite its importance, a fundamental trade-off remains between screening throughput and biological depth~\cite{gryniukova2023ai,dimasi2016innovation,lian2025inductive}. While chemical structures are readily available for billions of molecules, these static representations often fail to capture the dynamic functional landscapes of cellular responses~\cite{masarone2025advancing,ross2022large}. In contrast, high-content phenotypic profiling—such as transcriptomics and morphological imaging—provides direct readouts of drug action~\cite{chandrasekaran2024three, subramanian2017next,chandrasekaran2021image}. However, generating these profiles requires expensive wet-lab experiments, making it economically impractical to profile the vast chemical libraries accessible via virtual screening~\cite{haghighi2022high, seo2024pharmaconet, wang2023deepsa}. Bridging this gap by inferring functional signatures directly from molecular structures offers a transformative solution.

Two critical challenges impede progress toward this objective. First, the relationship between chemical structure and biological phenotype is highly non-linear and often decoupled; structurally distinct compounds can produce similar biological effects (functional synonymy), whereas minor chemical modifications may result in pronounced `activity cliffs.' Second, in contrast to the stability of chemical structures, biological readouts are inherently stochastic and subject to substantial batch effects and assay-specific artifacts~\cite{li2025phenoprofiler,schafer2024improving}. As a result, computational models that map structure to phenotype risk overfitting to these technical variations, failing to capture the transferable functional features necessary for accurate in silico profiling~\cite{he2021review}

To address these challenges, DECODE (DEcomposing Cellular Observations of Drug Effects) is introduced as a framework that distills latent biological semantics into structural representations. In contrast to fusion methods that require multi-modal inputs during inference, DECODE leverages paired transcriptomic and morphological data as privileged information available only during training. Through a geometric decoupling mechanism and contrastive alignment, the framework isolates a shared biological consensus from modality-specific noise. This approach enables the structural encoder to serve as a robust biological proxy, capable of inferring complex functional fingerprints solely from cost-effective molecular structures.

Our evaluations demonstrate that DECODE effectively bridges structural efficiency and biological accuracy. In zero-shot drug retrieval, it identifies functionally similar compounds even across high structural diversity. In sparse-label MoA prediction, it achieves a 20\% improvement over standard chemical baselines. More importantly, in external virtual screening campaigns for anti-cancer and target-specific agents, DECODE yielded a six-fold increase in hit rates for novel compounds. These results position DECODE as a powerful and scalable paradigm for biologically-informed virtual screening.
\begin{figure*}
    \centering
    \includegraphics[width=0.9\textwidth]{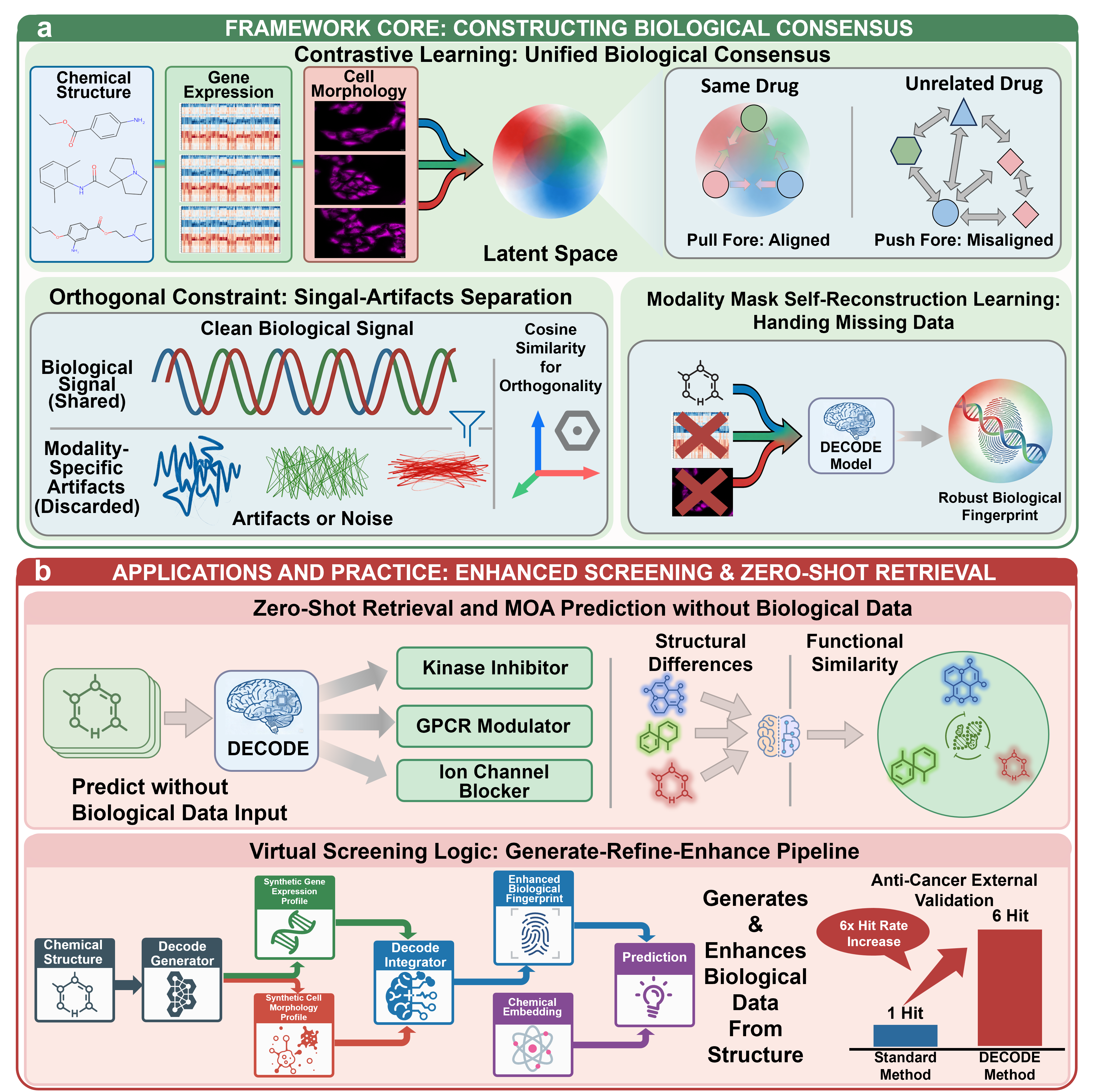}
    \caption{(a) The DECODE Framework for Modal Augmentation in Drug Discovery. Constructing a Unified Biological Consensus: The architecture integrates chemical structures with high-content transcriptomic and morphological profiles. It uses Contrastive Learning to align heterogeneous views into a shared latent space and Orthogonal Constraints to separate the measurement-invariant biological signal from modality-specific artifacts. A self-reconstruction task with modality masking ensures the learned fingerprint is robust to missing data. (b) Structure-Only Inference and Applications: The trained model enables high-fidelity in silico biological profiling using only chemical inputs. It supports Zero-Shot Retrieval, identifying functionally similar drugs despite structural diversity. For virtual screening, a `Generate-Refine-Enhance' pipeline integrates biological context, achieving a 6-fold increase in hit rates for novel active compounds compared to standard methods. }
    \label{fig:1}
\end{figure*}
\section{Methods}\label{sec2}
\subsection{Problem Formulation: Learning with Privileged Biological Context}

As depicted in Figure~\ref{fig:1}(a), we address the task of predicting drug bioactivity in a 'structure-only' inference setting while leveraging rich biological data available during training. Let $\mathcal{D} = \{(x_{d}^i, x_{g}^i, x_{m}^i, y^i)\}_{i=1}^{N}$ denote a multimodal dataset where $x_{d}$ represents the chemical structure (SMILES), and $x_{g} \in \mathbb{R}^{d_g}$ and $x_{m} \in \mathbb{R}^{d_m}$ denote transcriptomic and morphological features, respectively. Following the Learning Using Privileged Information paradigm, biological modalities ($x_{g}, x_{m}$) are available only during training to guide structural representation learning. Our objective is to learn a mapping $f_{\theta}(x_d)$ that projects the chemical structure into a measurement-invariant Biological Consensus Space, producing a biological fingerprint $z$ that captures the intrinsic effects of a drug independent of experimental noise.

\subsection{Dose-Aware Structural Encoding}

The chemical structure $x_{d}$ serves as the anchor for drug identity~\cite{zhang2022application}. MolFormer is utilized as the backbone $E_{struct}$ to map $x_{d}$ to a dense embedding $h_{d} \in \mathbb{R}^{h}$~\cite{ross2022large}. To account for the non-linear impact of dosage $x_{dose}$, a gating mechanism is introduced to modulate the structural features. The dose-specific gate $g$ and the dose-aware embedding $e_{d}$ are computed as follows:

\begin{equation}
    \begin{aligned}
g = \sigma(W_{g} \cdot E_{dose}(x_{dose}) + b_{g})
    \end{aligned}
\end{equation}
\begin{equation}
    \begin{aligned}
 e_{d} = \text{Encoder}_{drug}(h_{d} \odot g)
    \end{aligned}
\end{equation}

Here, $\sigma$ denotes the sigmoid function, $\odot$ indicates element-wise multiplication, $E_{dose}$ is a learnable dosage embedding, and $W_g$ and $b_g$ are learnable parameters. Modality-specific encoders concurrently process $x_{g}$ and $x_{m}$ into initial embeddings $e_{g}$ and $e_{m}$.

\subsection{Geometric Signal Disentanglement and Alignment}

To separate core biological signals from experimental artifacts, DECODE decomposes each modality embedding $e_k$ ($k \in \{d, g, m\}$) into two orthogonal subspaces: a shared consensus component $s_{k} = G_{s}(e_{k})$ and a modality-specific noise component $u_{k} = G_{u,k}(e_{k})$. This decomposition is enforced through an Orthogonality Constraint ($\mathcal{L}_{ortho}$), which minimizes the cosine similarity between shared and unique components:
\begin{equation}
    \begin{aligned}
\mathcal{L}_{ortho} = \sum_{k} | \text{sim}(s_{k}, u_{k}) | + \sum_{i \neq j} | \text{sim}(u_{i}, u_{j}) |
    \end{aligned}
\end{equation}

where $\text{sim}(\cdot, \cdot)$ denotes cosine similarity.
To ensure that the structural embedding $s_d$ accurately reflects biological reality, a Contrastive Alignment loss ($\mathcal{L}_{contrast}$) based on InfoNCE is employed:

\begin{equation}
    \begin{aligned}
\mathcal{L}_{contrast} = -\log \frac{\exp(\text{sim}(s_{d}, s_{g}) / \tau)}{\sum_{j=1}^{B} \exp(\text{sim}(s_{d}, s_{g}^j) / \tau)}
    \end{aligned}
\end{equation}

In this context, $\tau$ is the temperature parameter and $B$ is the batch size. This objective effectively transfers biological knowledge from the privileged modalities into the structural encoder.

The final functional fingerprint $z$ is generated by concatenating the centroids of the shared and unique features across the available modalities $V$:

\begin{equation}
    \begin{aligned}
  z = \text{Concat}(\bar{s}, \bar{u}), \quad \text{where } \bar{s} = \frac{1}{|V|}\sum_{k \in V} s_k, \quad \bar{u} = \frac{1}{|V|}\sum_{k \in V} u_k
    \end{aligned}
\end{equation}

This averaging mechanism ensures that the resulting fingerprint $z$ remains stable and biologically consistent, independent of the number of input modalities.
Model optimization is performed using a total objective $\mathcal{L}_{total} = \mathcal{L}_{recon} + \lambda_{1}\mathcal{L}_{ortho} + \lambda_{2}\mathcal{L}_{contrast}$, where $\mathcal{L}_{recon}$ denotes a Maximum Mean Discrepancy (MMD) loss that ensures feature reconstruction accuracy.

\subsection{Adaptive Inference Protocols}

As shown in Figure~\ref{fig:1}(b), DECODE implements three distinct inference strategies to accommodate varying data and task requirements.

Protocol I (Zero-Shot) infers $z$ from $x_d$ without requiring task-specific supervision, enabling applications such as mechanism of action (MOA) retrieval.

Protocol II (Dynamic Adaptation) robustly computes $z$ by leveraging any available subset of modalities $V$ through a masking strategy.

Protocol III (Generative Integration) employs a "Generate-Refine-Enhance" pipeline for virtual screening. A frozen generator $G_{frozen}$ synthesizes putative profiles $\hat{x}_{g}, \hat{x}_{m} = G_{frozen}(x_{d})$. Subsequently, a trainable integrator $F_{trainable}$ refines these profiles into a fingerprint $z_{bio} = F_{trainable}(x_{d}, \hat{x}_{g}, \hat{x}_{m})$. The final prediction is computed as $y_{pred} = \text{MLP}(\text{Concat}(z_{bio}, h_{d}))$.

\subsection{Data Collection and Benchmarking}
The foundation for manifold alignment was established using the  LINCS~\cite{natoli2021broadinstitute} and CDRP~\cite{bray2017dataset} datasets, providing paired observations of chemical perturbations with L1000 gene expression and Cell Painting morphological profiles. For downstream evaluation, we utilized the Non-Oncology Cancer Activity (NOCA) dataset~\cite{corsello2020discovering} for broad-spectrum anti-cancer prediction and the MedChemExpress Lung Cancer (MCELC) dataset for pathway-specific inference. Precision in ligand-based virtual screening was further validated on curated ChEMBL datasets for four targets: BACE1, COX-1, COX-2, and EP4.

We compared DECODE against robust baselines, including a Structure-Only baseline (MolFormer backbone~\cite{ross2022large}) and Bio-Only baselines trained on raw phenotypic data. This comparison validates whether our disentanglement mechanism can effectively overcome the technical artifacts and batch effects inherent in raw biological measurements. Detailed curation and implementation parameters are provided in Supplementary Information S1-S2.

\begin{figure*}[!ht]
    \centering
    \includegraphics[width=0.9\textwidth]{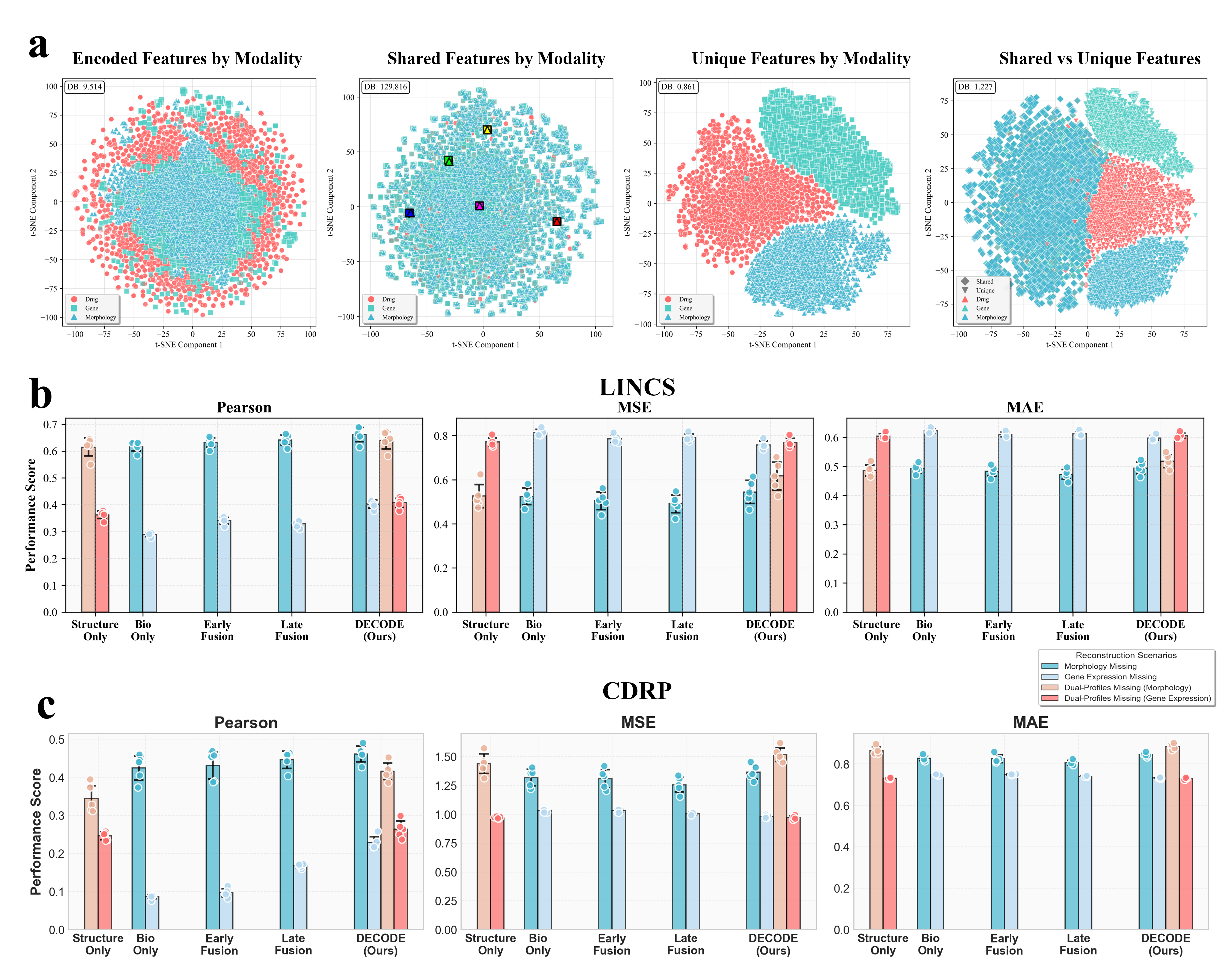}
    \caption{Geometric Analysis of Latent Disentanglement: t-SNE visualizations of the learned feature spaces. The plots reveal that the Shared Encoder successfully aligns heterogeneous modalities into a unified biological consensus (overlapping clusters in Shared Features), while the Orthogonal Constraints force modality-specific artifacts into distinct, non-overlapping subspaces (Unique Features), confirming effective signal purification.}
    \label{fig:2}
\end{figure*}

\section{Results}\label{sec3}

\subsection{Cross-Modal Reconstruction and Latent Space Analysis}
To determine whether DECODE internalizes the complex relationship between chemical structures and cellular responses, its performance was evaluated in cross-modal reconstruction and latent space organization.

Geometric Disentanglement of Signal and Noise. The central hypothesis underlying DECODE is that a robust biological fingerprint should be measurement-invariant. Visualization of the latent space using t-SNE demonstrates that features from different modalities (structure, morphology, and expression) corresponding to the same drug perturbation converge onto a compact, unified manifold (Figure.~\ref{fig:2}(a)). This convergence indicates that DECODE extracts a shared functional signal that transcends specific assay types. Additionally, the orthogonality constraint segregates modality-specific artifacts into distinct subspaces, ensuring that the structural encoder remains focused on core therapeutic effects rather than experimental noise.

High-Fidelity Phenotypic Inference. The model's ability to perform in silico biological profiling was further evaluated by reconstructing phenotypic signatures from chemical inputs. Across LINCS and CDRP datasets, DECODE consistently demonstrated superior reconstruction fidelity compared to single-view and standard fusion baselines (Figure~\ref{fig:2}(b)(c)). In cases where biological profiles were unavailable, the model generated morphological signatures that maintained high Pearson correlation with ground-truth physiological measurements. Qualitative analysis indicates that these generated profiles accurately reflect dose-dependent variations characteristic of specific MoA, such as BCL and SRC inhibitors (see Supplementary Section S3 for visual profiles and transcriptomic variance analysis).

\begin{figure*}
    \centering
    \includegraphics[width=0.9\textwidth]{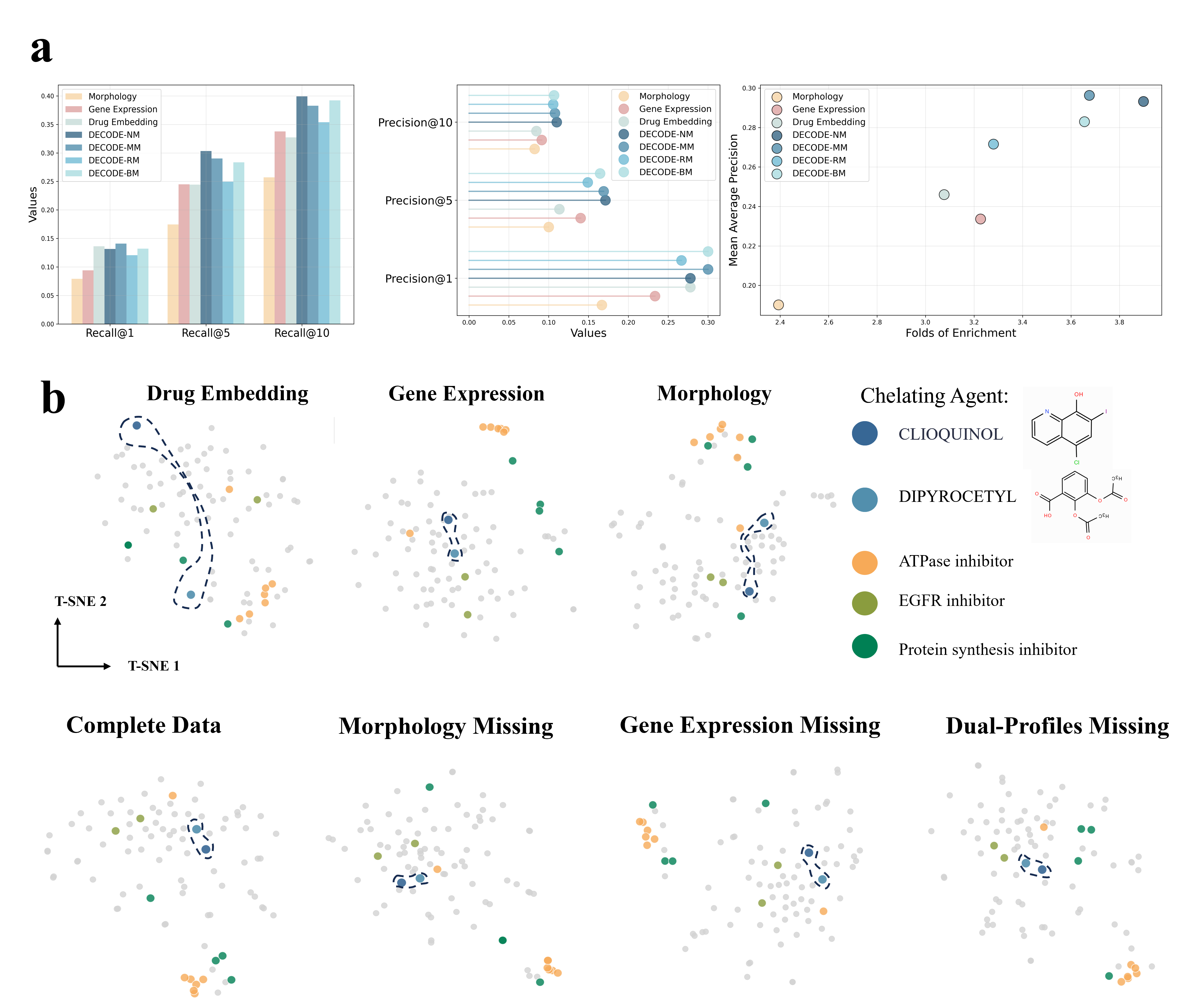}
    \caption{Zero-Shot Functional Retrieval and Generalization to Novel Chemical Spaces. Quantitative Retrieval Evaluation: Comparative analysis of retrieval metrics (Recall, Precision, and Mean Average Precision vs. Enrichment) in  Novel Chemical Space. The DECODE-BM variant (Dual-Profiles Missing, or Structure-Only Inference) consistently outperforms single-modality baselines, showing the model's ability to generalize biological insights to unseen chemical entities without wet-lab data. (b) Visualizing Functional Alignment: t-SNE projections highlight the `Chelating Agent' class (e.g., Clioquinol and Dipyrocetyl). Despite significant structural dissimilarity, as shown by the dispersed Drug Embedding, DECODE's biological fingerprint clusters these functionally related drugs. This semantic grouping remains robust even in the Dual-Profiles Missing scenario, confirming that the model has disentangled the shared Mechanism of Action signal from structural and experimental variations.}
    \label{fig:3}
\end{figure*}

\subsection{Zero-Shot Retrieval of Functional Synonyms (Protocol I)}
After validating the alignment mechanism, Protocol I was implemented to assess whether the structural encoder enables zero-shot functional retrieval in the absence of task-specific supervision.

Capturing Functional Synonyms. This protocol assesses the model's capacity to identify drugs with identical Mechanisms of Action (MoAs) using only their chemical structures ($x_d$). In a rigorous evaluation within novel chemical space, defined as testing on non-duplicate compounds, DECODE's structure-based fingerprint substantially outperformed baselines that utilize raw biological modalities or standard chemical fingerprints . DECODE achieved up to over 20\% relative improvement in Top-5 Recall compared to morphology method (Figure~\ref{fig:3}(a)).

Generalization Beyond Structural Similarity. DECODE addresses the limitations of traditional structural similarity approaches. For instance, chelating agents such as CLIOQUINOL and DIPYRO-CETYL, which have distinct chemical scaffolds, are mapped to a tight cluster in DECODE's functional latent space despite their structural differences. In contrast, for heterogeneous classes like protein synthesis inhibitors, the model maintains natural biological dispersion rather than enforcing artificial clustering (Figure~\ref{fig:3}(b)). These findings indicate that the structural encoder has progressed from recognizing molecular syntax to capturing `functional synonyms,' thereby enabling high-fidelity zero-shot inference of drug mechanisms.

\begin{figure*}
    \centering
    \includegraphics[width=\textwidth]{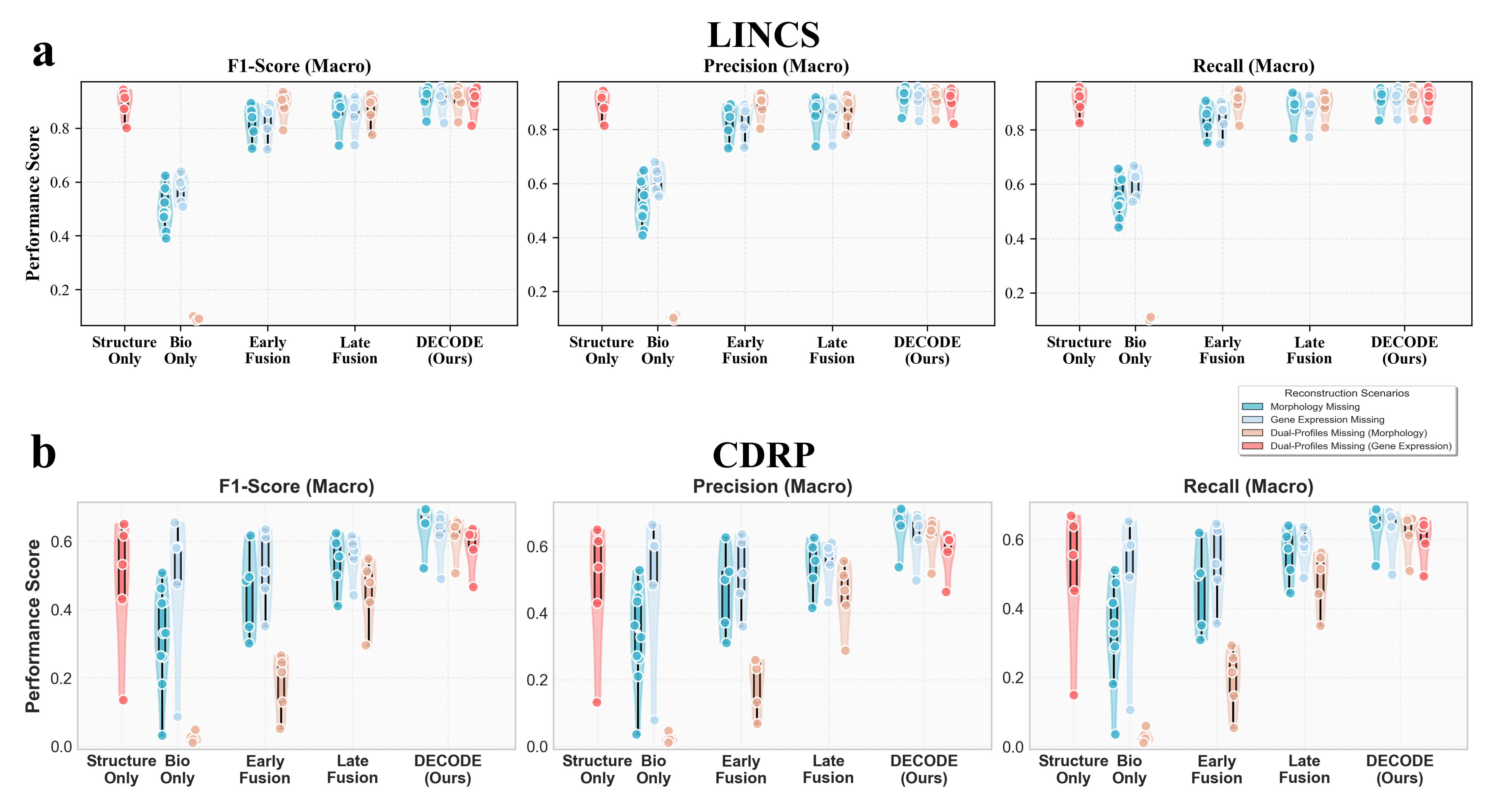}
    \caption{Comparative Mechanism of Action (MOA) Prediction and Geometric Disentanglement Analysis. Performance is compared using Macro F1-Score, Precision, and Recall across the LINCS (a) and CDRP (b) datasets. DECODE consistently outperforms Single-View (Structure or Bio Only) and standard Fusion baselines (Early or Late Fusion).}
    \label{fig:4}
\end{figure*}

\subsection{Robustness in MOA Prediction (Protocol II)  }

We evaluated DECODE's robustness in supervised MoA classification across dynamically varying data modalities. On the sparsely labeled CDRP dataset, DECODE achieved a 15.8\% relative improvement in F1-score over the Expert MLP baseline (Figure~\ref{fig:4}(a)).

A key finding is DECODE's capacity to mitigate noise accumulation. In standard Late Fusion baselines, incorporating transcriptomic data with morphological inputs reduced performance (F1-score decreased from 0.554 to 0.537), indicating that raw biological data may introduce conflicting noise. In contrast, DECODE's performance improved under identical conditions (F1-score increased from 0.619 to 0.642). This result indicates that the disentanglement mechanism functions as a dynamic filter, selectively integrating consensus biological signals and discarding modality-specific artifacts. Similar trends were observed in the LINCS dataset (Figure~\ref{fig:4}(b)), and feature distribution analysis confirmed that the consensus space remains stable after fine-tuning (see Supplementary Section S4).

\begin{figure*}[ht]
    \centering
    \includegraphics[width=\textwidth]{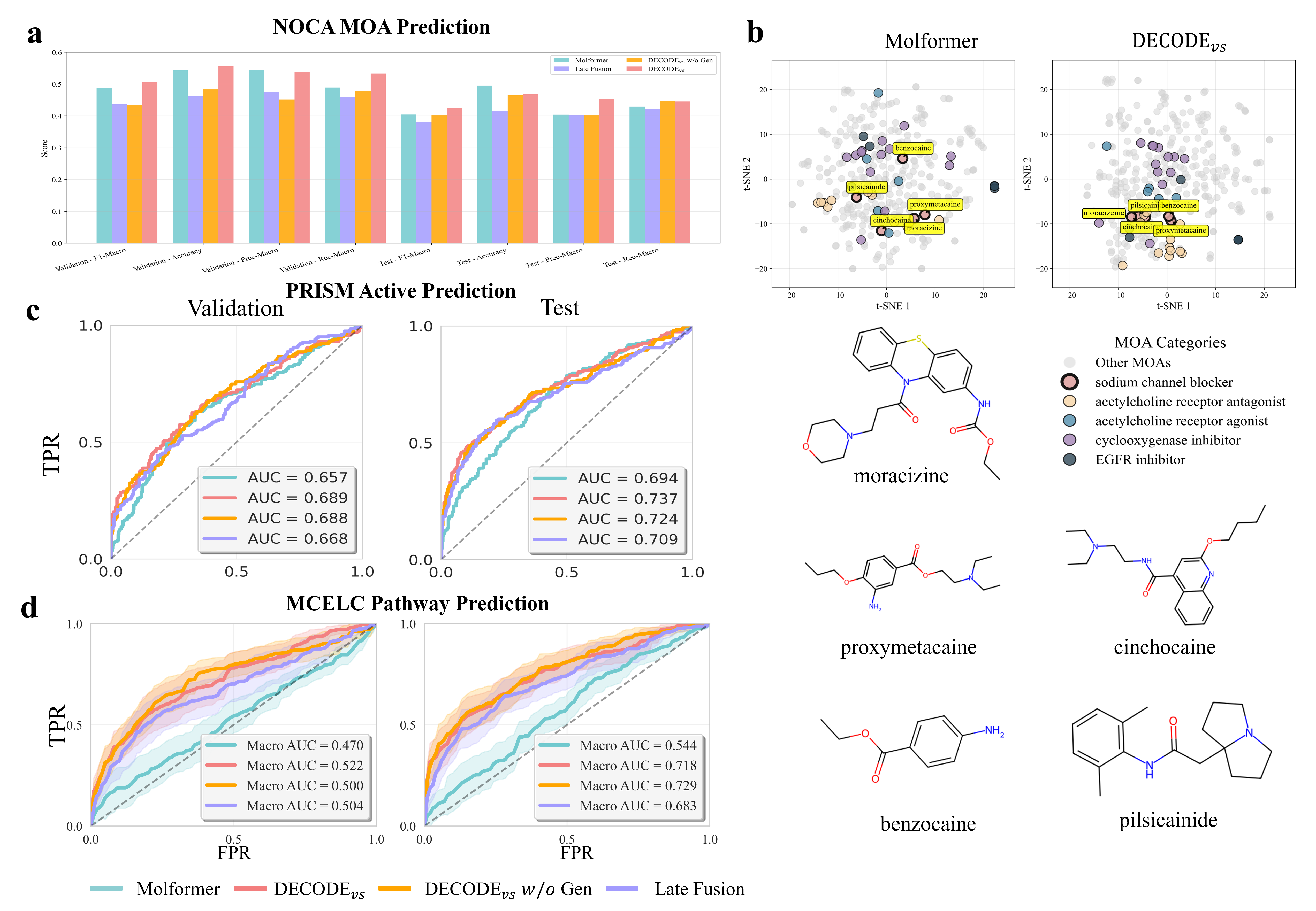}
    \caption{(a), DECODE outperforms a structure-only model (Molformer) in MOA prediction on the NOCA dataset.
(b), t-SNE visualization reveals that DECODE learns a more coherent latent space, grouping functionally related drugs, such as the highlighted sodium channel blockers, more effectively.
(c), In external anti-cancer screening, DECODE achieves a higher AUC than the chemical baseline.
(d) The structural visualization of sodium channel blockers.DECODE achieves a higher macro-AUC in predicting drug pathways on the MCELC dataset. The shaded area represents the 95$\%$ confidence interval derived from the distribution of AUCs across all pathway classes. }
    \label{fig:5}
\end{figure*}
\subsection{Augmenting Virtual Screening (Protocol III)}
To address the `structure-only' challenge in virtual screening (VS), a 'Generate-Refine-Enhance' pipeline was implemented. This protocol employs DECODE as a generator to synthesize putative biological profiles for novel compounds.

Superiority in Out-of-Distribution (OOD) Tasks. On the Non-Oncology Cancer Activity (NOCA) dataset, the full $DECODE_{vs}$ pipeline achieved the highest accuracy (Macro F1 0.442), outperforming both the MolFormer baseline and a direct transfer strategy without generative augmentation (Figure~\ref{fig:5}(a)). These results indicate that, in novel biological contexts, generated profiles provide essential privileged information that bridges the domain gap. This generative bridge is demonstrated by DECODE's ability to link structurally diverse sodium channel blockers, such as moracizine and proxymetacaine, which chemical-only models do not cluster (Figure~\ref{fig:5}(b), MoA clustering analysis).

External Validation and Hit Rate Improvement. In external anti-cancer activity screening, $DECODE_{vs}$ achieved an AUC of 0.737, which is significantly higher than the chemical baseline (0.694) (Figure~\ref{fig:5}(c)). Notably, $DECODE_{vs}$ identified six confirmed active hits among the top-ranked candidates, representing a six-fold increase in hit rate compared to MolFormer (Figure~\ref{fig:5}(d)).

Task-Specific Adaptation. This evaluation was extended to the MedChemExpress Lung Cancer (MCELC) dataset and four target-specific tasks (BACE1, COX-1, COX-2, EP4). In familiar biological contexts, such as lung cancer cell lines, direct transfer of the structure encoder was sufficient to achieve high performance (Macro AUC 0.729), particularly for cytoskeleton-related pathways (Supplementary Section S5). In contrast, for novel targets or broad phenotypic screens, generative augmentation remained the optimal strategy. Across all ligand-binding tasks, biologically-informed representations consistently improved hit rates over standard chemical embeddings (see Supplementary Section S5).

\subsection{Ablation Studies: Dissecting the Mechanism of Disentanglement}
A systematic ablation study on the LINCS and CDRP datasets confirms the necessity of each architectural component in DECODE (detailed metrics in Supplementary Section S6). Our analysis reveals that the modal alignment phase is the most critical factor for performance; removing this phase (w/o Modal Alignment) leads to a substantial decline in F1-score on the CDRP dataset (from 0.642 to 0.589). Furthermore, a naive 'Joint' training strategy—which attempts simultaneous reconstruction and classification without explicit disentanglement—yielded even lower performance (0.555). This suggests that aligning chemical representations with a biological consensus prior to downstream tuning is essential for extracting robust signals from high-dimensional, noisy data.

The importance of the geometric constraints is particularly evident in the ``structure-only'' inference setting. Eliminating the contrastive alignment loss resulted in the most pronounced degradation in retrieval performance (LINCS F1-score falling from 0.584 to 0.529), indicating that without explicit contrastive guidance, the chemical encoder fails to internalize the biological manifold. Similarly, removing the orthogonality constraint reduced the F1-score to 0.552, supporting our conclusion that enforcing geometric separation between shared consensus signals and modality-specific artifacts is vital for learning a measurement-invariant biological fingerprint. Overall, the synergy between privileged biological supervision and geometric disentanglement is fundamental to DECODE's capacity for biologically-informed discovery.
\section{Conclusion}\label{sec4}
This work introduces DECODE, a framework that shifts the computational drug discovery paradigm from purely structure-based screening to structure-based in silico biological profiling. By leveraging multi-modal transcriptomic and morphological data as privileged information during training, DECODE distills deep functional semantics into cost-effective chemical representations. This enables the generation of high-fidelity biological fingerprints solely from molecular structures during inference, bridging the gap between the scalability of virtual screening and the biological depth of phenotypic assays.

Empirical evidence demonstrates that DECODE captures the intrinsic MoA with high accuracy, even in zero-shot scenarios. Most notably, in external anti-cancer screening, DECODE achieved a six-fold improvement in hit rates over standard baselines, highlighting its utility as a distillation engine that transforms noisy biological signals into actionable therapeutic insights.

While DECODE establishes a robust foundation for learning biological function from structure, future work will focus on two key areas. First, we aim to incorporate context-aware injection mechanisms to account for drug-response heterogeneity across diverse tissue types, extending beyond the current single-cell line context. Second, we plan to integrate biological foundation models as feature extractors to further enhance the resolution of the learned consensus space. Ultimately, DECODE paves the way for a more efficient and biologically-informed era of drug discovery by democratizing access to complex phenotypic insights.

\medskip
\textbf{Acknowledgements} \par 
This work was supported by the National Natural Science Foundation of China (Grant No. 62432011, 62450002).
\bibliography{reference}
\bibliographystyle{oup-plain}
\end{document}